# Stable Perovskite Solar Cells via exfoliated graphite as an ion diffusion-blocking layer

Abdullah S. Alharbi[1,+], Miqad S. Albishi[1,+], Temur Maksudov[2], Tariq F. Alhuwaymel[1], Chrysa Aivalioti[2], Kadi S. AlShebl[1], Naif R. Alshamrani[1], Furkan H. Isikgor[2], Mubarak Aldosari[1], Majed M. Aljomah[1], Konstantinos Petridis,[3] Thomas D. Anthopoulos[2,4], George Kakavelakis[3, *], Essa A. Alharbi[1, 2 *]

[1] Microelectronics and Semiconductor Institute, King Abdulaziz City for Science and Technology (KACST), Riyadh 11442, Saudi Arabia

[2] KAUST Solar Center (KSC), King Abdullah University of Science and Technology (KAUST), Thuwal 23955-6900, Saudi Arabia

[3] Department of Electronic Engineering, School of Engineering, Hellenic Mediterranean University, Romanou 3, Chalepa, Chania, Crete GR-73100, Greece

[4] Henry Royce Institute and Photon Science Institute, Department of Electrical and Electronic Engineering, The University of Manchester, Manchester M13 9PL, UK

[+] These authors contributed equally to this work.
**\* Authors to whom correspondence should be addressed: kakavelakis@hmu.gr and ealharbi@kacst.edu.sa**

## Abstract

Ion and metal diffusion in metal halide perovskites, charge-transporting layers, and electrodes are detrimental to the performance and stability of perovskite-based photovoltaic devices. As a result, there is an intense research interest in developing novel defect and ion diffusion mitigation strategies. We present a simple, low-cost, scalable, and highly effective method that uses spray-coated exfoliated graphite interlayers to block ion and metal diffusion and humidity ingress within the perovskite, the hole transport material, and metal electrodes. The influence of inserting the exfoliated graphite films on the structural, surface morphology, and optoelectronic properties were examined through various methods, including X-ray diffraction, Time-of-Flight Secondary Ion Mass Spectrometry, Scanning electron microscope, atomic force microscopy, Current-voltage (*J-V*) characteristics, Transient photocurrent, and transient photovoltage. Our comprehensive investigation found that exfoliated graphite films reduced the $I^-$ and $Li^+$ diffusion among the layers, leading to defect mitigation, reducing non-radiative recombination, and enhancing the device stability. Consequently, the best-performing device demonstrated a power conversion efficiency of 25% and a fill factor exceeding 80%. Additionally, these devices were subjected to different lifetime tests, which significantly enhanced the operational stability.

# Introduction

Perovskite solar cells (PSCs) have recently emerged as a groundbreaking photovoltaic (PV) technology in pursuit of efficient and sustainable energy sources. PSCs power conversion efficiencies (PCEs) soar above 26%,[1] combined with cost-effective and low temperature (<150 ºC) fabrication processing, making them a promising candidate for a sustainable transition to clean and renewable energy[2,3]. The crucial factors for successfully translating PV technology from laboratory to commercial products are low cost, efficiency, stability, and scalability.[4] These metrics should be considered alongside other factors, such as low toxicity and energy payback times. Hybrid metal halide perovskites are highly efficient PV materials that have garnered significant interest in fundamental and applied research. However, their practical significance still needs to be determined due to the limited device operational lifetimes. The inherent instability of PSC components is a significant challenge for their practical use.[5] The main issue is the sensitivity of the organic halide in perovskite to moisture, which restricts its long-term stability under humid conditions.[5–7] External encapsulation was suggested to obtain highly stable and efficient PSCs to protect the perovskite layer from ambient humidity and thermal stress and suppress the nonradiative recombination.[8–11] In particular, the most common configuration used in PSCs is the *n-i-p* configuration, which contains LiTFSI and tBP-doped Spiro-OMeTAD hole transporting layer (HTL) between the gold electrode and the perovskite absorber.[12–14] However, performance degradation is still present in this type of configuration, even after encapsulation or in a moisture-free environment.[15] One of the main reasons for the poor stability in the presence of Spiro-OMeTAD HTL in *n-i-p* PSCs' device configuration is due to the lithium ions (Li$^+$) diffusion towards the perovskite layer that creates an undesirable path of degradation at the perovskite/HTL interface and the perovskite film itself as well as causing the ion migration.[16] Another problem is the unwanted reaction between iodine ions (I$^-$) and gold (Au), causing degradation of the gold electrode and the metal/HTL interface [17]. This reaction is exacerbated by positive bias under constant light exposure, leading to a significant irreversible drop in PCE.[18–20] Another fundamental degradation pathway is the diffusion of gold atoms (mainly at elevated temperatures) from the metal electrode towards the perovskite layer, which induces degradation and performance loss.[21] Effective strategies have been widely implemented to solve these issues between the Spiro-based HTL and the perovskite layer. Since most methods are based on organic or inorganic salts (such as OAI, PEAI, DMAI, EAI, CsI, RbI, etc.),[22–24] they still proceed by a spin-coating deposition method, which does not provide an up-scalable target PSCs, which is crucial for the future large-scale deployment of PSCs. To solve these problems, only a few materials have been implemented successfully via scalable methods: 1) evaporated thin films such as $MoO_3$,[25] Cr, and $CrO_2$[26], which are not desirable due to the time-consuming and cost-intensive extra thermal evaporation step; and 2) the use of reduced graphene oxide (rGO) but only in low PCE (~20%) CuSCN-based PSCs,[27] as rGO cannot be used for the high PCE PSCs based on Spiro HTL, due to its solubility in the same solvent (chlorobenzene) as Spiro-OMeTAD (rGO washes out Spiro HTL).

Graphene, a carbon allotrope with a hexagonal lattice nanostructure, has gained attention in the scientific community due to its intriguing characteristics, such as high conductivity and transparency.[28] It can be quickly processed at large scales and low temperatures using liquid-phase exfoliation methods.[29–31] Various graphene-related materials (GRMs), including single-layer graphene, multi-layer graphene, nanoflakes, nanoribbons, and nanoplatelets, have been extensively studied and applied in PSCs.[32–35] Incorporating GRMs into PSCs has shown promising results in cost-effectiveness and improved performance.[36] GRMs are also thermally and environmentally stable, hydrophobic, and mechanically robust. The ultra-small pore size of graphene flakes makes them suitable for filtering or blocking tiny atoms and gases. An interesting approach relevant to graphene's properties and has not been demonstrated yet is its effectiveness as a passive blocking layer for metal and ion diffusion in PSCs to enhance their stability against moisture-induced degradation.[37] To date, most of the methods employed in the literature to prepare exfoliated graphite inks applied in PSC result in flakes that are dispersed in boiling point solvents that are not desirable and compatible with halide perovskites.[35] That is why GRMs have not been successfully employed on highly efficient PSCs as protection-blocking layers to improve their operational stability.

Herein, we report the incorporation of scalable double-side ion diffusion-blocking layers between the perovskite/HTL and the HTL/Au interfaces in PSCs using exfoliated graphite films produced by rotor-stator mixer homogenization of graphite in ethanol. Our approach to disperse the exfoliated graphite powders in a low boiling point solvent (ethanol) and combine its deposition with a scalable method known as spray coating (instead of spin-coating) makes exfoliated graphite ink compatible with deposition on the sensitive surfaces of perovskite and Spiro-OMeTAD films. Thus, our strategy to simultaneously protect the perovskite and Spiro-OMeTAD surfaces with the hydrophobic and robust exfoliated graphite film significantly increased the operational lifetime of the optimized devices with exfoliated graphite barriers. Also, the optimized devices presented a slight improvement in the PCE compared to the exfoliated graphite-free PSCs. Finally, we performed a detailed analysis including X-ray diffraction (XRD), Time-of-Flight Secondary Ion Mass Spectrometry (TOF-SIMS), scanning electron microscope (SEM), atomic force microscopy (AFM), current-voltage (J-V) characteristics, Transient photocurrent (TPC) and transient photovoltage (TPV) decay measurements, to uncover the mechanism causing this improvement in the stability and PCE (from 24.33% to 25.01%), upon the introduction of exfoliated graphite between the two interfaces (perovskite/HTL and HTL/Au).

## Experimental

*Rotor stator homogenization production of EGF*: The homogenization of graphite ink in NMP and DCB (~50 g/L) was performed using a 4-blade rotor (32 mm) placed in a fixed stator (rotor–stator gap ~100 µm). In a typical RS homogenization experiment, 0.5L of graphite ink was processed at 4500 rpm for 60 minutes. After the homogenization process, the resultant exfoliated graphite flakes (EGF) dispersions were centrifuged (10000 rpm for 30 minutes) to remove the remaining unexfoliated graphite flakes. Afterwards, the upper 70% of the supernatant was collected, and the EGF powder was collected after drying the homogenized ink's solvent (NMP and DCB). To prepare the EGF inks used as ion diffusion-blocking layers in perovskite solar cells, different amounts of the EGF powder were dispersed in ethanol (0.05 – 0.25 mg/ml) as the dispersion medium.

*Characterization techniques for EGF*: The thickness of the EGF was measured with a Profilometer from Bruker (Dektak XT Stylus). The SEM images of the EGF films were measured using a ZEISS Merlin HR-SEM. The lateral size distribution of the EGFs was obtained by analyzing several SEM images of flakes deposited on glass. The Raman spectra of the EGF films were obtained using a Horiba

Scientific LabRAM-HR Raman microscope (514 nm excitation laser with a 50x lens and 5 spectra averaged).

*Device Fabrication*: Fluorine-doped tin oxide (FTO)-glass substrates (1.8mm) were etched and cleaned by ultrasonication in Hellmanex (2%, deionized water), rinsed thoroughly with de-ionized water and ethanol, and then treated in oxygen plasma for 30 min. Approximately 25 to 30 nm of blocking layer (TiO2) was sprayed on the cleaned FTO at 450 °C using a commercial titanium diisopropoxide bis(acetylacetonate) solution (75% in 2-propanol, Sigma-Aldrich) diluted in anhydrous ethanol (1:6 volume ratio). A 150 nm mesoporous TiO2 layer (diluted paste (1:7 wt. ratio of Dyesol 50NRD: ethanol)) was applied by spin-coating at 5000 rpm for 15 s and then sintered at 450 °C for 30 min in dry air. The perovskite films were deposited using a single-step deposition method from the precursor solution which was prepared in an Argon atmosphere and contained 1.7 M of FAI MABr, PbBr2, and PbI2 with an added 30–35 mol% of MACl in anhydrous dimethylformamide/dimethylsulphoxide (4:1 (volume ratio)) to achieve the desired composition: $FA_{0.98}MA_{0.02}Pb(I_{0.98}Br_{0.02})_3$ (3% PbI2 excess). Perovskite solution was spin-coated in a two-step program at 1000, followed by 5000 rpm. Subsequently, 270 μl of chlorobenzene was dropped on the spinning substrate. This was followed by annealing the films at 150 °C for 35–40 min. Different concentrations of EGF diluted in ethanol were prepared and sprayed on the cleaned perovskite film or/and Spiro-Ometad at various temperatures for surface treatment via EGF. Device fabrication was carried out inside a dry air box under controlled atmospheric conditions. To complete device fabrication, 92 mg of 2,2',7,7'-tetrakis(N, N-di-p-methoxyphenylamine)-9,9-spirobifluorene (spiro-OMeTAD) were dissolved in 1 ml of chlorobenzene to be used as hole-transporting material (HTM) and doped with bis(trifluoromethylsulfonyl)imide lithium salt (24 μl of a solution prepared by dissolving 520 mg LiTFSI in 1.0 ml of acetonitrile), and 40 μl of 4-tert-butylpyridine. The PTAA was dissolved in chlorobenzene (12 mg/ml) with additives of 7.5 μl bis(trifluoromethane)sulphonimide lithium salt/acetonitrile (170 mg ml$^{-1}$) and four μl tBP. Both HTMs were spin-coated at 5000 rpm for 30 s. Finally, a ~120 nm gold (Au) layer was applied via thermal evaporation.

*Device Characterization*: The J–V curves were measured at ambient temperature and air conditions using a Keithley 2400 source unit under simulated AM 1.5 G solar illumination at 100 mW cm−2 (1 sun). The light intensity was calibrated through a KG-3 Si diode with a solar simulator (Enli Tech, Taiwan). The devices were recorded in reverse and forward scans in the range of 0–1.2 V with voltage steps of 10 mV and a settling time of 100 ms and 0–1.4 V. The photo-active area of 0.158 cm$^2$ was defined using a dark-colored metal mask. A polymer film stuck to the front glass was used as an antireflection coating. Transient photovoltage (TPV), transient photocurrent (TPC), were measured by the all-in-one characterization system (PAIOS, Fluxim, Switzerland).

*Long-Term Light Soaking Test*: The devices were measured using a maximum power point (MPP) tracking routine under continuous illumination at room temperature with a sample holder purged with continuous nitrogen flow. The MPP was updated, and a Reverse scanned J–V curve was recorded every 40 minutes.

*Scanning Electron Microscopy (SEM)*: SEM analysis was performed on a ZEISS Merlin HR-SEM.

*Powder X-ray diffraction (XRD)*: The Powder X-ray diffractograms were recorded using an ARL EQUINOX 1000 X-ray Diffractometer.

*Atomic force microscopy (AFM)*: The surface morphology was studied using the Bruker dimension ICON AFM system.

*Time-of-Flight Secondary Ion Mass Spectrometry (TOP-SIMS)*: The Hiden TOF-SIMS system is used for the devices' elemental identification and depth profiling.

## Results and Discussion

Figure S1 illustrates a schematic of the rotor-stator mixer (RSM) used to induce high shear in the graphite-solvent precursor dispersion through a closely spaced rotor/stator combination.[38] In this process, graphite flakes are first dispersed in a 'good' solvent[38], such as dichlorobenzene (DCB) and N-Methyl-2-pyrrolidone (NMP). Afterward, the application of the RMS to the precursor dispersion induces the exfoliation and fragmentation of graphite flakes (due to the presence of high shear rates) and, thus, the production of few-layer graphene or, more precisely, exfoliated graphite flakes (EGF). The optimized conditions for the RSM process were developed in another work [27] and are discussed in detail in the supporting information. After processing the graphite precursor with the RSM process, the dispersion became more homogeneous and stable over time (see Figure 1a for graphite precursor compared to the EGF ink). This improved stability of the inks after RSM processing is due to their exfoliation and fragmentation, i.e., the flakes become significantly smaller and thinner and thus are substantially lighter compared to their graphite counterparts. Therefore, the RSM processing of the precursor dispersions resulted in more stable dispersions that can be processed more efficiently, which is favourable for their application/printing on functional surfaces. This is vital for their use and adaptation in functional devices such as solar cells.

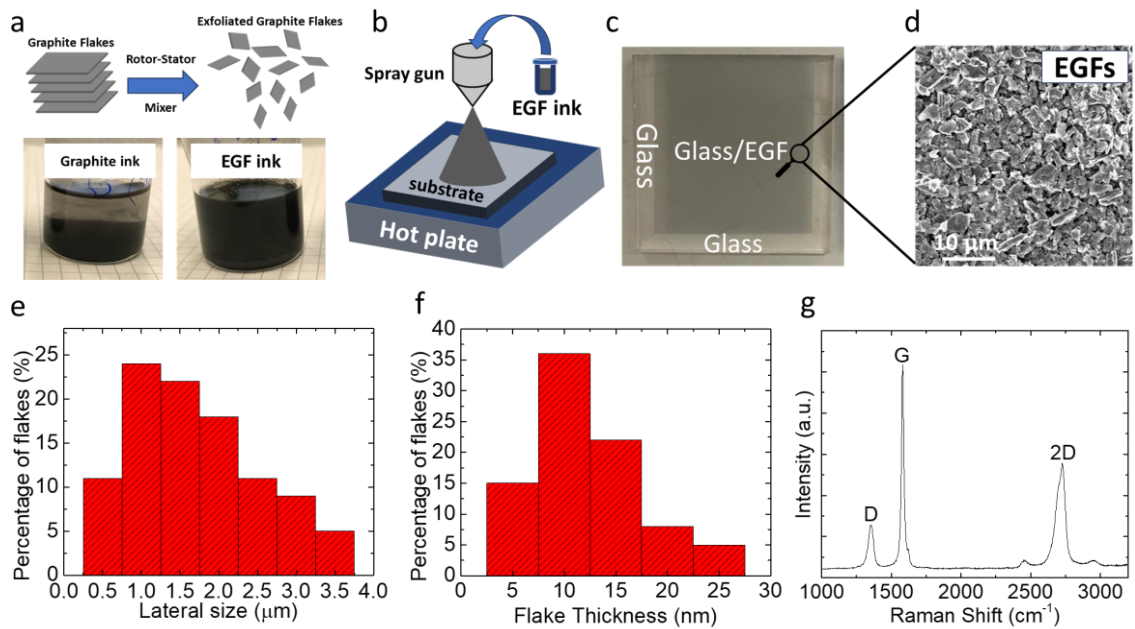

**Figure 1**. a) Schematic illustration and images of the graphite flakes and inks before and after the RMS process application to the precursor dispersion. b) A schematic illustration of the EGFs-based film deposition using spray coating, c) an image of an as-deposited EGF-based film patterned using a shadow mask, d) SEM image of the EGF-based film, e) the flakes size distribution, f) the flake thickness distribution and g) the Raman spectroscopy of the EGF-based film.

Graphene-based coatings, due to their chemical robustness and environmental stability[31] and their hydrophobic nature[39], are promising candidates to improve the lifetime of chemical and environmentally unstable devices. Among the different deposition/printing methods for fabricating graphene-based coatings from dispersions, spray-coating is among the most reliable techniques for fabricating uniform, reproducible, and continuous films over large areas.[39,40] We also tested and optimized the deposition of our EGF ink dispersions (redispersed in ethanol at 0.1 mg/ml EGF) using spray-coating as the deposition method. We also evaluated the spin-coating deposition method, a typical solution-processed deposition process. Our findings suggest that spin-coating results in non-homogeneous and irreproducible films (Figure S2). That is why we also moved our focus to spray-coating. As shown in the schematic of Figure 1b, as-prepared EGF-based ink (in ethanol) was inserted into a manually controlled spray gun container and then sprayed on the target substrates using controlled spray pressure and nozzle-to-substrate distance.[31] It is essential to mention here that the spray-coating procedure described above was conducted while heating the deposition substrate (at a temperature that was close to the boiling point of the solvent medium of the ink) to allow the quick removal of the solvent, which turned to be critical for the uniform films to be obtained.

Following the procedure, homogeneous coatings were reproducibly obtained over relatively large areas (such as the one in Figure 1c 2.5 cm x 2.5 cm), especially considering the flake size of a single EGF (~ 1 μm in lateral dimensions). Another benefit of this deposition approach for our EGF-based inks is that specific patterns with sharp edges can, on demand, be obtained by simply using a specifically designed shadow mask. Several SEM images and Raman spectra were collected and analysed to assess the uniformity of the EGF-based coatings. As shown in the SEM image of Figure 1d, uniform coating over relatively large areas and relatively smooth film morphology was obtained by multiple EGFs overlapping each other, forming a compact film. This excellent substrate coverage and uniformity were obtained by optimizing the spray-coating parameters (pressure, distance, and ink concentration). Also, the 2D nature of the EGFs facilitated the homogeneous formation of the respective films. It is worth mentioning that the deposition of the untreated graphite precursor ink through spray-coating was practically impossible since the flakes before exfoliation were extensive in the lateral dimension (with some flakes even exceeding 20 μm), often causing clogging of the spray nozzle. In Figures 1e and 1f, we also present the morphological characteristics of the EGFs. In Figure 1e, we show the EGF flake size distribution after applying the optimized RSM exfoliation process conditions (see supporting information). We observed that the highest percentage of the EGFs are approximately 1 μm long, and the vast majority (~75%) are below 2 μm. This implies that the RSM process significantly reduces the flake size of the graphite after exfoliation. In Figure 1f, we present the flake thickness of the optimized condition processed EGFs, where the highest percentage of them is approximately ~10 nm and more than 55% of them are below 10 nm. This implies that the RSM process reduces the flake size and effectively exfoliates graphite to few-layer graphene (or EGF, as mentioned above).

Besides the effect of EGF on exfoliation, we also assessed the quality of EGFs. Thus, Raman spectroscopy was also used to characterize the EGF-based compact films prepared using the spray-coating deposition method. First, in Figure 1g, we observed the characteristic sharp G peak at ~1581 cm$^{-1}$, whose origin is related to the in-plane vibrations of the graphene lattice. Upon exfoliation, we also observe an additional peak (compared to typical unexfoliated graphite), the so-called D peak at ~1352 cm$^{-1}$, generated in the presence of in-plane vibrations of the graphene lattice and originates from defects. The G peak ($I_G$) intensity is ~4 times that of the D peak ($I_D$). An important metric to assess the quality of EGF-based film is the $I_D/I_G$ ratio; in the case of EGF-based film, this ratio is ~0.25. This low $I_D/I_G$ ratio is only present in high-quality graphene-based flakes and samples.[31] Furthermore, all EGF-based films presented a strong 2D peak at ~2728 cm$^{-1}$, originating from a process where momentum conservation is satisfied by two phonons with opposite wave vectors.[31] The 2D peak is always present in graphene-based flakes and samples, as defects do not need to activate it.

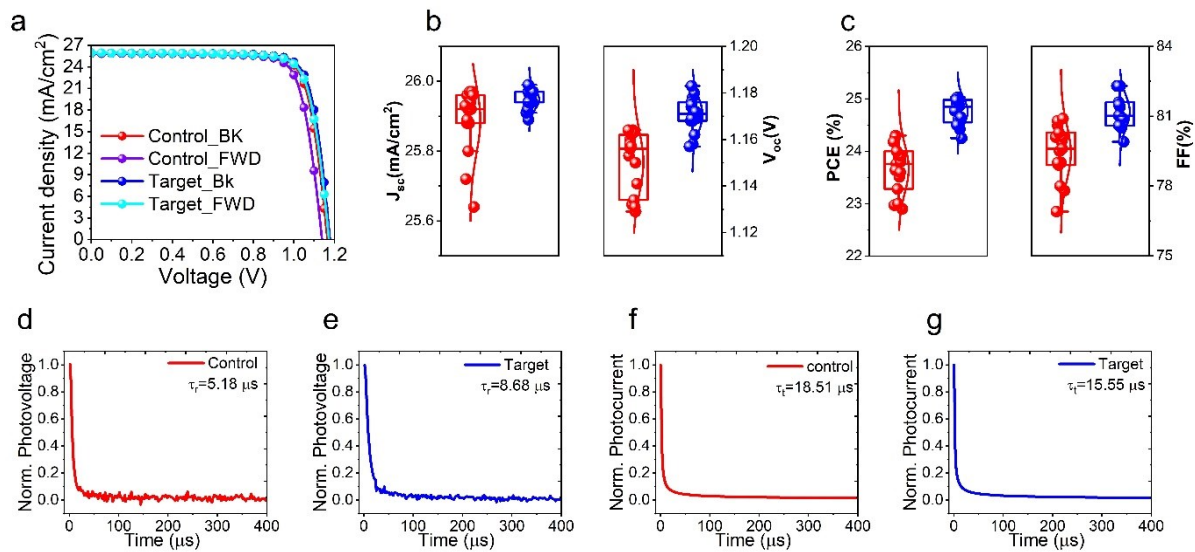

**Figure 2. Optoelectronic properties:** (a) *J–V* curves of the champion control and target solar cells under reverse and forward scan, b) and c) Summary of the photovoltaic metrics ($V_{oc}$, $J_{sc}$, PCE, and FF) of the devices (Red:

control, blue: target) d-e) Normalized transient photovoltage of control and target devices. f-g) Normalized transient photocurrent of control and target devices, respectively.

To investigate the influence of the EGF interlayer on the PSCs, a perovskite precursor solution with FAPbI$_3$ rich composition of (FAPbI$_3$)$_{0.98}$(MAPbBr$_3$)$_{0.02}$, (FA$^+$ = formamidinium = CH(NH$_2$)$_2^+$; MA$^+$ = methylammonium = CH$_3$NH$_3^+$) was used and the perovskite films were deposited by spin-coating onto the mesoporous TiO$_2$ (mp-TiO$_2$) layer by a one-step method using chlorobenzene as antisolvent. To begin with, we systemically optimized the double ion diffusion blocking-layer deposition of EGF (Device structure: Glass/FTO/c-TiO$_2$/m-TiO$_2$/Perovskite/EGF/Spiro-OMeTAD/EGF/Au) in terms of the ethanol/EGF ratio and preparation temperature as shown in Figure S3 and S4, Table S1 and S2. Hereafter, the pristine perovskite device (i.e. without EGF) is referred to as "control," and the perovskite device with EGF interlayers is referred to as the "target." The *J-V* curves of control and target devices are illustrated in Figure 2a. The target device achieved a PCE of 25.0% with an open-circuit voltage ($V_{oc}$) of 1.178 V, a fill factor (FF) of 81.6%, and a short-circuit current density ($J_{sc}$) of 25.98 mA/cm$^2$ in the backward scan. In the forward scan, the device had a PCE of 24.51 % with a $V_{oc}$ of 1.171 V, an FF of 80.7%, and a $J_{sc}$ of 25.96 mA/cm$^2$. On the other hand, the control devices presented a PCE of 24.33% ($J_{sc}$: 25.96 mA/cm$^2$, $V_{oc}$: 1.163 V, FF: 80.6%) in the backward scan and a PCE of 23.30% ($J_{sc}$: 25.88 mA/cm$^2$, $V_{oc}$: 1.132 V, FF: 79.5%) in the forward scan. This led to a hysteresis index (HI = [(PCE$_{backward}$ − PCE$_{forward}$)/(PCE$_{backward}$)]x100 of 2.23% and 4.15% for targeting the control devices, respectively. Moreover, the statistical distribution of the PV characteristics ($J_{SC}$, $V_{OC}$, FF, and PCE) is presented in Figures 2b and 2c, which shows improvement mainly in FF and $V_{oc}$ PV device parameters. This improvement in $V_{oc}$ and FF implies a reduction in non-radiative recombination and an improvement in charge carrier extraction in the presence of EGF-blocking layers in the device configuration. Furthermore, to confirm the validity and accuracy of our J-V measurements and the proper calibration of our solar simulator, we collected and calculated the external quantum efficiency (EQE) spectra and integrated Jsc (Figure S5) of the control and target devices. As it can be seen, the control and target PSCs presented comparable and of similar pattern EQE spectra with almost unchanged integrated $J_{SC}$ (which confirms our JV data presented in Fig. 2a). Importantly, we also noted a less than 4% difference for the Jsc of both devices compared to the values we measured in JV, confirming the validity and accuracy of our JV measurements. In the following parts of this manuscript, we will investigate the exact mechanisms responsible for the improvements mainly in FF and $V_{oc}$ of our target devices.

To gain further insights into the charge recombination and extraction dynamics, we performed the transient photocurrent (TPC) and transient photovoltage (TPV) decay measurements to investigate the effect of inserting the EGF on control and target devices. TPC and TPV decay measurements are frequently employed to ascertain PV devices' charge transport time ($\tau_t$) and charge recombination time ($\tau_r$).[41–43]. The photovoltage (PV) decay measured under open-circuit conditions offers valuable information about the lifespan of charge carriers. As shown in Figures 2d and e, the $\tau_r$ of the control devices is merely 5.18 μs, which is considerably shorter than the 8.68 μs observed in the target devices; thus, we observe a more than 65% increase in $\tau_r$ after inserting the EGF. This TPV result justifies our *J-V* measurement observation of improved $V_{oc}$ for the target devices. TPC decay experiments, conducted with pulsed light in a short-circuit state, enable the transportation and extraction of photo-generated carriers by the charge extraction layers ($TiO_2$ and Spiro-OMEDAT). Figures 2f and g indicate that control devices have a longer $\tau_t$ (18.51 μs), ~ 20%, than the target (15.55 μs) devices. This is a clear indication that a slightly faster charge extraction is taking place in the devices incorporating EGF in the device architecture. The TPC results are in full agreement with our observation in the improved FF values for the target samples compared to the control samples in the J-V measurements. The increased recombination time and shortened transport time are solid indications for reduced incidence of non-radiative recombination and carrier extraction enhancement in devices built using EGF-barrier layers.

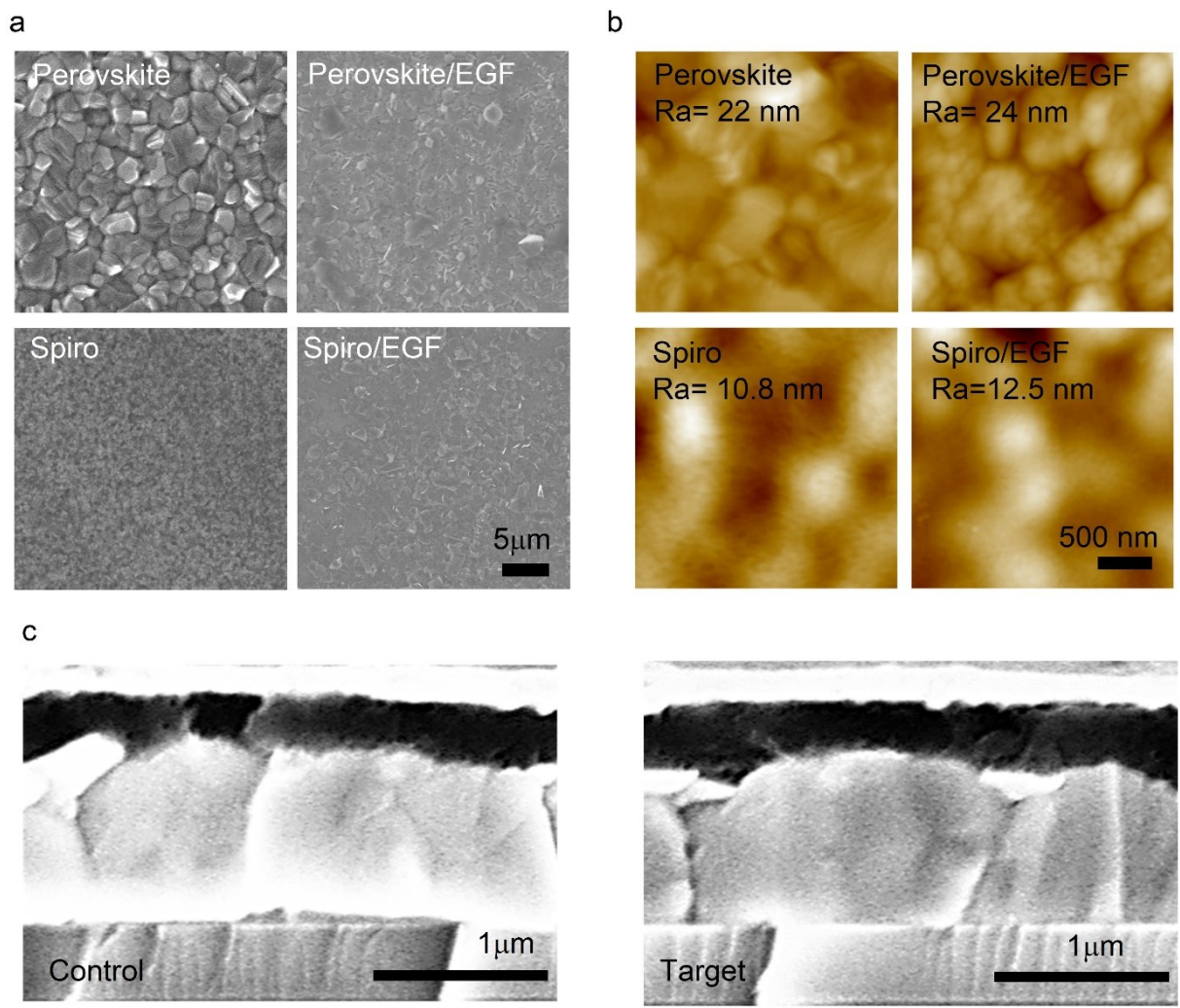

**Figure 3. Surface morphology:** a) and b) top-view SEM images and AFM images, respectively, of control-1 (perovskite) and control-2 (Spiro), along with corresponding target-1 and target-2 films incorporating overlying EGF films. c) shows a cross-section SEM of the control and target samples.

We performed further morphological, structural, and surface characterizations to gain insights on the effect of EGF on the morphology of the perovskite and Spiro-OMeTAD films. Furthermore, the SEM and AFM recorded the surface morphology and roughness images of Perovskite, Perovskite/EGF, Spiro, and Spiro/EGF films (Figures 3a and b), respectively. The top-view SEM image of the EGF-free perovskite film shows a uniform and compact morphology with visible grain boundaries. In contrast, after depositing the EGF on the perovskite surface, the grain boundaries are covered with a uniform EGF coating (Figure 3a). This means that a compact EGF layer was deposited and formed on top of the perovskite and Spiro films, respectively, which could protect them from moisture and harmful particles (such as ions or metals) diffusion. We also obtained AFM images of the mentioned above samples (Figure 3b). In particular, perovskite, perovskite/EGF, Spiro, and Spiro/EGF films presented the following surface roughness (Ra) values of 22 nm, 24 nm, 10.8 nm, and 12.5 nm, respectively. This suggests that the incorporation of EGF on perovskite and Spiro has almost no effect on their surface roughness because of EGF's extremely thin and uniform flake nature. Additionally, cross-section SEM images of the control and target devices were obtained (Figure 3c) to check the durability of the bulk PSC structure upon the deposition of the EGF layers. The EGF layer thickness in the optimized target devices was minimal and practically invisible from the cross-section SEM images; thus, we do not observe any noticeable difference in the cross-section SEM images. Additionally, this proves that EGF deposition does not affect the bulk of the perovskite layer or the whole device structure, which aligns with our previously obtained results from J-V, TPV, and TPC measurements.

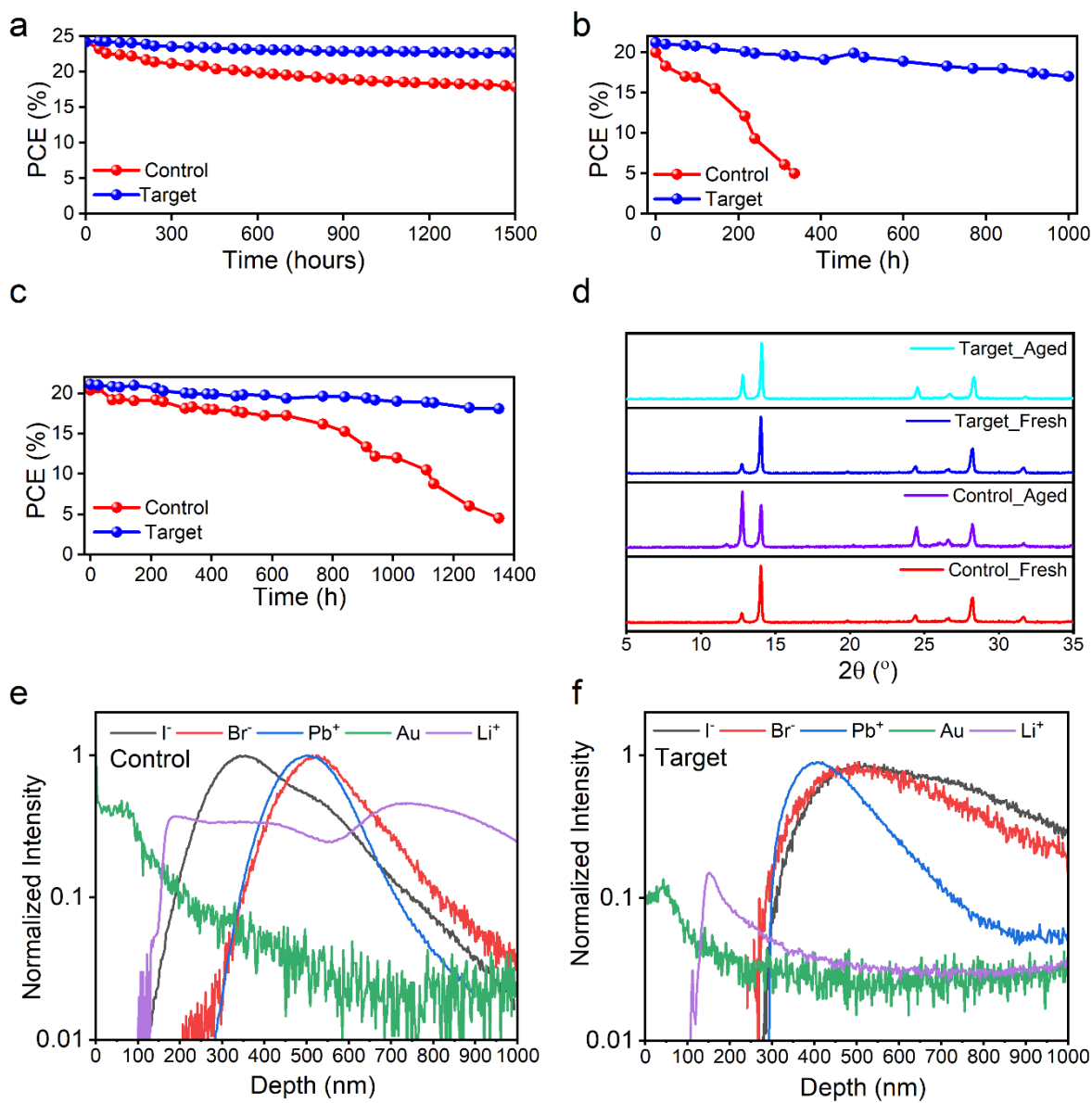

**Figure 4. Stability studies:** Evolution of the PCE of (a) control and target using Spiro as HTL under one sun illumination in an $N_2$ environment, (b) control and target devices under 1 Sun illumination at 80°C and 75% RH using PTAA as HTL (c) control and target devices stored in the dark at 80°C and 75% RH using PTAA as HTL, d) XRD pattern of perovskite films of fresh and aged, control and target samples. E) and f) ToF-SIMS of perovskite devices of control and target, respectively.

Besides efficiency, another important aspect of our work is related to the devices' operational lifetimes, which is the key challenge in the research field of PSCs. To evaluate and understand EGF's effect on our devices' lifetime, we applied different stability protocols on devices with and without EGF-treatment. Firstly, we performed operational stability by subjecting the control and target devices under one sun illumination at maximum power point (MPP) in a nitrogen environment at room temperature (RT). The control and target devices retained ~70% and ~90% of the initial performance after 1500 hours of continuous testing, respectively (Figure 4a). Thus, according to the results mentioned above, the insertion of EGF improved the operational lifetime by 20% compared to the control devices (EGF-free). This is an important result since MPP stability at full one-sun illumination is an important accelerating test that solar cells must surpass by at least 90% of the initial PCE after 1000 hours. Also, shelf-life stability was performed by keeping the control and target devices in the dark at RT and ~8% relative humidity (RH). This stability protocol is important to check if the materials introduced in the device structure are self and fundamentally stable and whether there is an interaction between the materials without an accelerating factor. By analyzing the results, our control and target devices retained ~ 90% and ~97% of the initial performance after 1600 hours, respectively (Figure S6). This means both devices (control and target) are quite stable without inserting an accelerating factor such as temperature, humidity, bias, etc. However, we observe a small improvement of ~7% for target devices, which means there is a small intrinsic instability for control devices even without an accelerating factor. Additionally, we observed that this intrinsic instability of control devices is almost fully suppressed with the insertion of EFG. That is why we decided to perform a demanding accelerating lifetime test to check and identify the troublemaker for the intrinsic stability and the effect of EGF on the stability. Thus, we stressed the control and target devices at 80 °C and 75% RH under one sun illumination. By using this test, the acceleration factor is increased significantly, and we can test the real stability of our devices and whether the troublemaker material can be a feasible option for the commercialization of PSC technology. We found that the control and the target devices based on the Spiro-OMeTAD hole transporter degraded significantly after 120 hours (Figure S7). Of course, the insertion of EGF significantly improved (by ~50%) the stability of these unstable control devices. Still, even these control devices cannot be considered an option for the commercial landscape. Thus, we concluded that Spiro-OMeTAD is the troublemaker for stable PSCs, and that is why we decided to focus on alternative HTLs to test EGF's effect on our PSCs' lifetime. It is well known that HTL plays a crucial role in the stability of PSCs. Specifically, Spiro- OMeTAD, when used as an HTL, has been previously reported to be unstable at temperatures above 60 °C and high humidity[26]. Hence, we decided to focus on PTAA HTL (see Figure S8and Table S3) instead of Spiro-OMeTAD, as it is more stable and tolerant to temperature and humidity for the stability test at 80 °C and 75% RH. This allowed us to analyze EGF's effect on the PSC's stability further when exposed to harsh conditions. The target device (with PTAA HTL and EGF) achieved a PCE of 21.77%, $V_{oc}$ of 1.142 V, FF of 74.4%, and $J_{sc}$ of 25.62 mA/cm$^2$ in the backward scan and a PCE of 21.04% $V_{oc}$ of 1.121 V, FF of 73.3%, and $J_{sc}$ of, 25.61 mA/cm$^2$ in the forward scan. On the other hand, the control device presented a PCE of 20.86 % ($J_{sc}$: 25.41 mA/cm$^2$, $V_{oc}$: 1.132 V, FF: 72.5%) in the backward scan and a PCE of 17.31% ($J_{sc}$: 25.39 mA/cm$^2$, $V_{oc}$: 1.07 V, FF: 63.61%) in the forward scan which reflected a hysteresis index (HI = [(PCE$_{backward}$ − PCE$_{forward}$)/(PCE$_{backward}$)]x100 of 3.35% and 17.01% for target the and control devices, respectively. We can see that EGF treatment enhances performance and significantly reduces hysteresis. We then performed detailed stability experiments at different conditions using the more stable PTAA as HTL using the more stable PTAA as HTL at different conditions. The shelf-life stability of the control, perovskite/EGF/HTL, perovskite / HTL/EGF, and target devices was tested in the dark at room temperature and approximately 8% relative humidity, as shown in Figure S9. The target device (with double EGF barrier layers) retained 92% of its initial value, significantly better than the control device, which only retained 25% of its initial value. Afterwards, in Figures 4b and c we present the accelerated stability test conducted under illumination and dark at 80°C under 75% RH without encapsulation, respectively. The control devices retained 15% of the initial value during a 350-hour test under one-sun illumination and 57% after 1000 hours in dark conditions. Conversely, the target devices showed excellent retention of the initial PCE, maintaining over 90% of the initial performance after 1000 hours in both scenarios. This striking stability improvement with the addition of EGF strongly indicates its function to stabilize and protect the interfaces and sensitive materials of the device structure.

Thus, to get an insight into the role of EGF, we aged our films at 80°C and 75% RH for 75 hours and recorded the XRD patterns of the fresh and aged control and target films, as presented in Figure 4d. XRD is a powerful tool for studying and providing information about materials' structure and chemical composition. Our strategy to check the XRD patterns before and after exposing the control and target films is to identify the potential changes in the perovskite structural and chemical composition and whether EGF insertion affects it. We can see that the PbI$_2$ peak after aging becomes more pronounced in the control film at 12.6$^0$ in the XRD pattern, and there is an emergence of the δ-FAPbI$_3$ phase peak at 11.5$^0$ in the XRD pattern only of the aged control samples. This indicates that the control samples' degradation process involves structural and composition changes. On the other hand, the fresh and aged target films remained stable and almost unchanged after 75 hours without any δ-FAPbI$_3$ phase[44] in the XRD pattern. The incorporation of EGF stabilized the perovskite film's structure and chemical composition. Consequently, we concluded that EGF treatment significantly enhanced the perovskite solar cell stability because it stabilized the perovskite layer, as we saw no effect in its XRD pattern with the incorporation of EGF.

To further support our essential findings from the XRD measurements, we also conducted a ToF-SIMS study. TOF-SIMS is a powerful analytical technique sensitive to all components of PSCs, which can provide insight into the structural integrity, the phase of the material intermixing, and the sharpness of the interfaces within the devices. This study also allowed for observing changes in the chemical composition of the devices' depth profile. To obtain all this vital information, TOF-SIM was conducted for the control and target devices after aging them at 80°C and 75% RH for 100 hours to understand the effect of inserting the EGF film on the properties mentioned above (Figure 4e, f). After aging the samples, several elements such as Gold (Au), Bromide (Br$^-$), Lead (Pb$^{2+}$), Iodide (I), and Lithium (Li$^+$) were tracked in both control and target devices. Both devices show similar depth profiles in terms of Br$^-$ and Pb$^{2+}$. However, differences are observed in the depth of I$^-$ and Li$^+$. In particular, in the control samples (Figure 4 e), Li$^+$ diffused from Spiro-OMeTAD towards the perovskite layer, and I$^-$ diffused from the perovskite layer into Spiro-OMeTAD. In contrast, no ion or metal diffuses between the layers in the target device, as shown in Figure 4 f. This indicates that the EGF films serve as a barrier in the target devices, preventing the ion migrations of I$^-$ and Li$^+$ and resulting in less hysteresis and outstanding stability performance. Thus, EGF barriers kept our devices' interfaces very sharp and avoided the diffusion of ions and metals within the device. This is directly linked with the very stable XRD patterns of the target samples, where we observe no changes indicating no change in the perovskite composition and structure. The stability of the perovskite is explained by the EGF protecting it from harmful ions and metal particles. On the other hand, the severe changes in the XRD patterns of control samples are also explained by the apparent diffusion of ions and metal particles we observed in TOF-Sims that affected the perovskite structure and chemical composition after aging. This characteristic of the EGF film is directly linked and explains the increased devices' stability and durability, making them more suitable for long-term use. In conclusion, the TOF-SIMS results further supported our findings from the XRD patterns and stability studies of control and target samples, making the indication very concrete and specific.

**Conclusions**

We successfully utilized EGF films in PSCs to enhance their long-term operational stability. The EGF, prepared through rotor-stator mixer homogenization, was found to act as a diffusion barrier layer for metal particles and ions at the perovskite/HTL and HTL/Au interfaces. Solar cells with the EGF barrier layer showed increased PCE (from 24.33% for control to 25% for target) and extended operational lifetime. Microstructural analysis using XRD of the perovskite in cells with and without the EGF barrier before and after aging showed that the $PbI_2$ peak (12.60°) was more prominent in the control sample (without EGF) than in the target sample (with EGF), indicating improved structural stability of the perovskite in the presence of the EGF layer. Additionally, TOF-SIMS analysis revealed that EGF hindered the diffusion of metal and ions between the layers, resulting in reduced hysteresis and enhanced cell stability due to the improved structural integrity of the perovskite layer. The proposed EGF barrier layer technology offers a cost-effective, scalable, and straightforward approach to enhancing the long-term operational stability of PSCs. We believe this approach could be applicable to other metal-halide perovskite-based devices, such as perovskite-based light-emitting diodes, photodetectors, transistors, gas sensors, etc. Our findings provide valuable new insights that can help expedite the commercialization of PSCs while positively impacting other device technologies.

**Conflicts of interest**

There are no conflicts to declare

**Data availability**

The data supporting this article have been included as part of the Supplementary Information

**Acknowledgments**

G.K. and K.P. gratefully acknowledges the support from the Hellenic Mediterranean University and the Department of Electronics Engineering. E.A.A gratefully acknowledges the support of King Abdulaziz

City for Science and Technology (KACST), Saudi Arabia and and the Ibn Rushd Fellowship, King Abdullah University of Science and Technology (KAUST). A.S.A , M.A, T.F.A and N.R.A acknowledges the support of King Abdulaziz City for Science and Technology (KACST), Saudi Arabia.

## Author Contributions

E.A.A and G.K conceived the idea of the work, designed, planned the experiments and supervised the work. A.S.A, M.S.A., T.F.A, K.A, and M.A fabricated and optimized the perovskite solar cell devices, did all the basic characterizations, analyzed the data and wrote the manuscript with support from G.K. and E.A.A. G.K. T.M, C.A and F.H.I were responsible for SEM, RAMAN, TPV and TPC measurement and analysis. A.S.A and M.M.A performed the AFM and Tof-SIM measurements, respectively. T.D.A. and K.P. revised the initial draft of the manuscript. All authors contributed towards the preparation of the manuscript and approved its submission.

## Notes

The article has been submitted for publication to Journal of Materials Chemistry A. After it is published, it will be found at https://pubs.rsc.org/en/journals/journalissues/ta#!recentarticles&adv.

## Bibliographic references

# Supporting Information

**Stable Perovskite Solar Cells via exfoliated graphite as an ion diffusion-blocking layer**

Abdullah S. Alharbi[1+], Miqad S. Albishi[1+], Temur Maksudov[2], Tariq F. Alhuwaymel[1], Chrysa Aivalioti[2], Kadi S. AlShebl[1], Naif R. Alshamrani[1], Furkan H. Isikgor[2], Mubarak Aldosari[1], Majed M. Aljomah[1], Konstantinos Petridis,[3] Thomas D. Anthopoulos[2,4], George Kakavelakis[3, *], Essa A. Alharbi[1, 2 *]

[1] Microelectronics and Semiconductor Institute, King Abdulaziz City for Science and Technology (KACST), Riyadh 11442, Saudi Arabia

[2] KAUST Solar Center (KSC), King Abdullah University of Science and Technology (KAUST), Thuwal 23955-6900, Saudi Arabia

[3] Department of Electronic Engineering, School of Engineering, Hellenic Mediterranean University, Romanou 3, Chalepa, Chania, Crete GR-73100, Greece

[4] Henry Royce Institute and Photon Science Institute, Department of Electrical and Electronic Engineering, The University of Manchester, Manchester M13 9PL, UK

[+] These authors contributed equally to this work
**\* Authors to whom correspondence should be addressed: kakavelakis@hmu.gr and ealharbi@kacst.edu.sa**

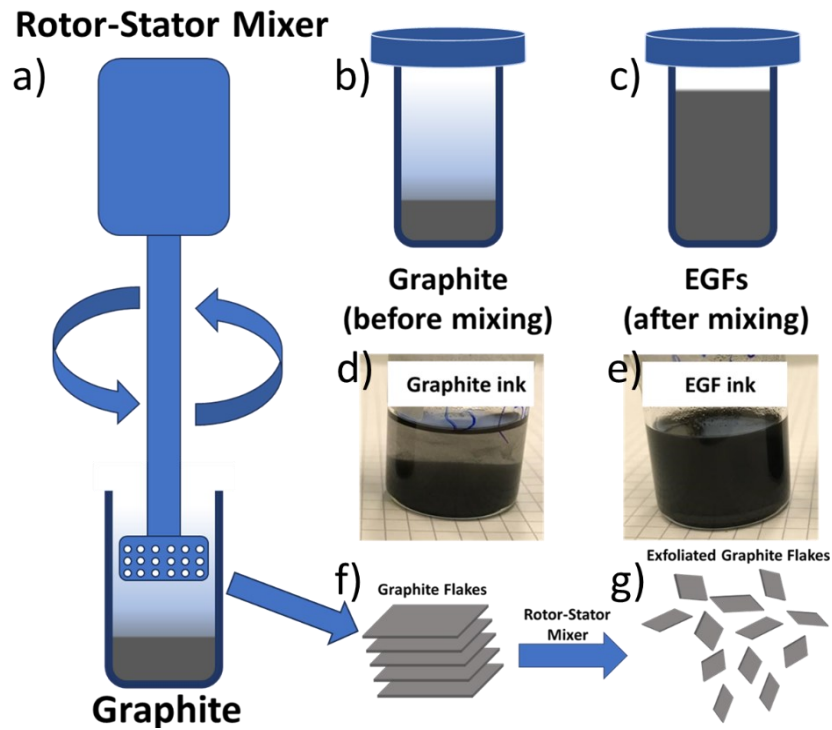

**Figure S1**. (a) Schematic illustration of RSM exfoliation/fragmentation process. Schematic illustration and images of the graphite flakes and inks (b,d,f) before and (c,e,g) after the RMS process application to the precursor dispersion.

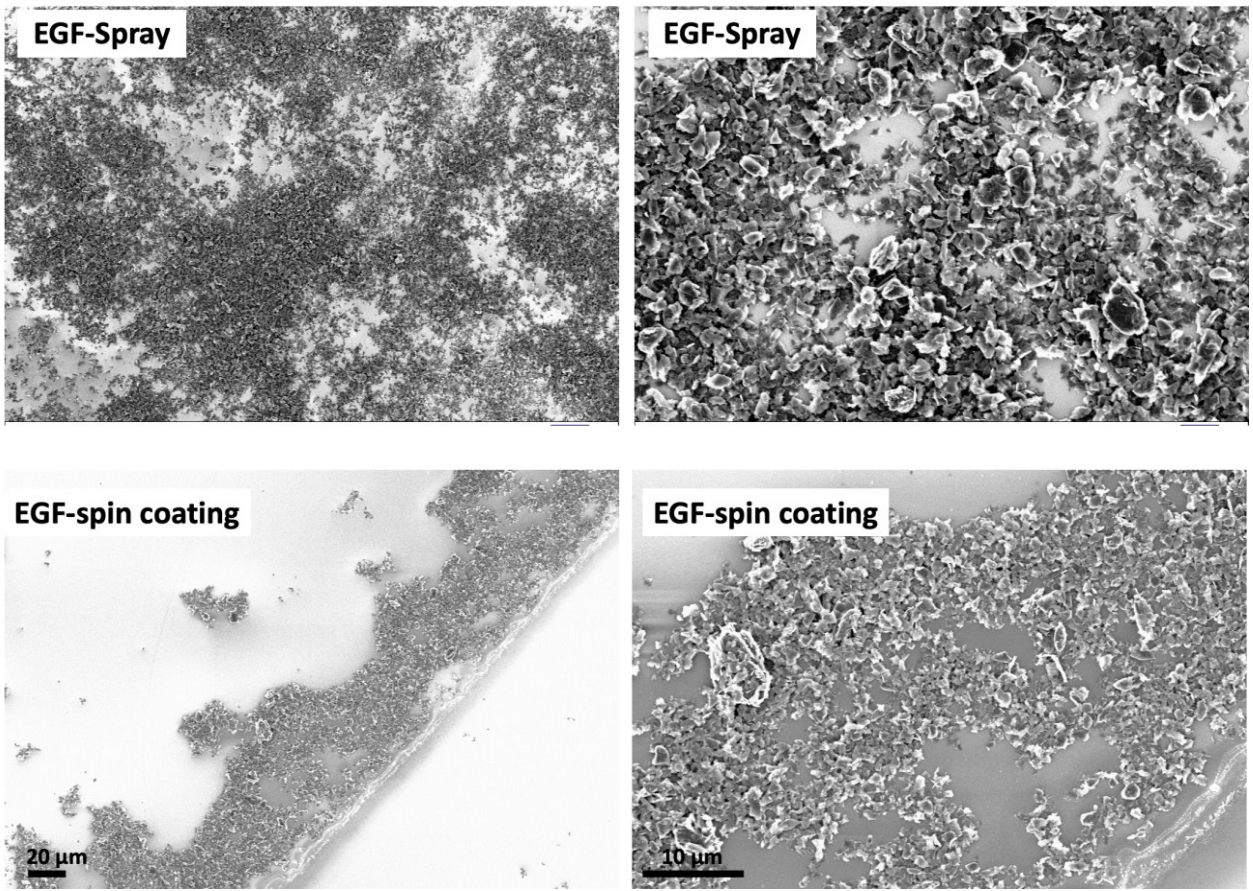

**Figure S2.** Top-view SEM images comparing the surface coverage of graphene flakes deposited through spray- and spin-coating techniques.

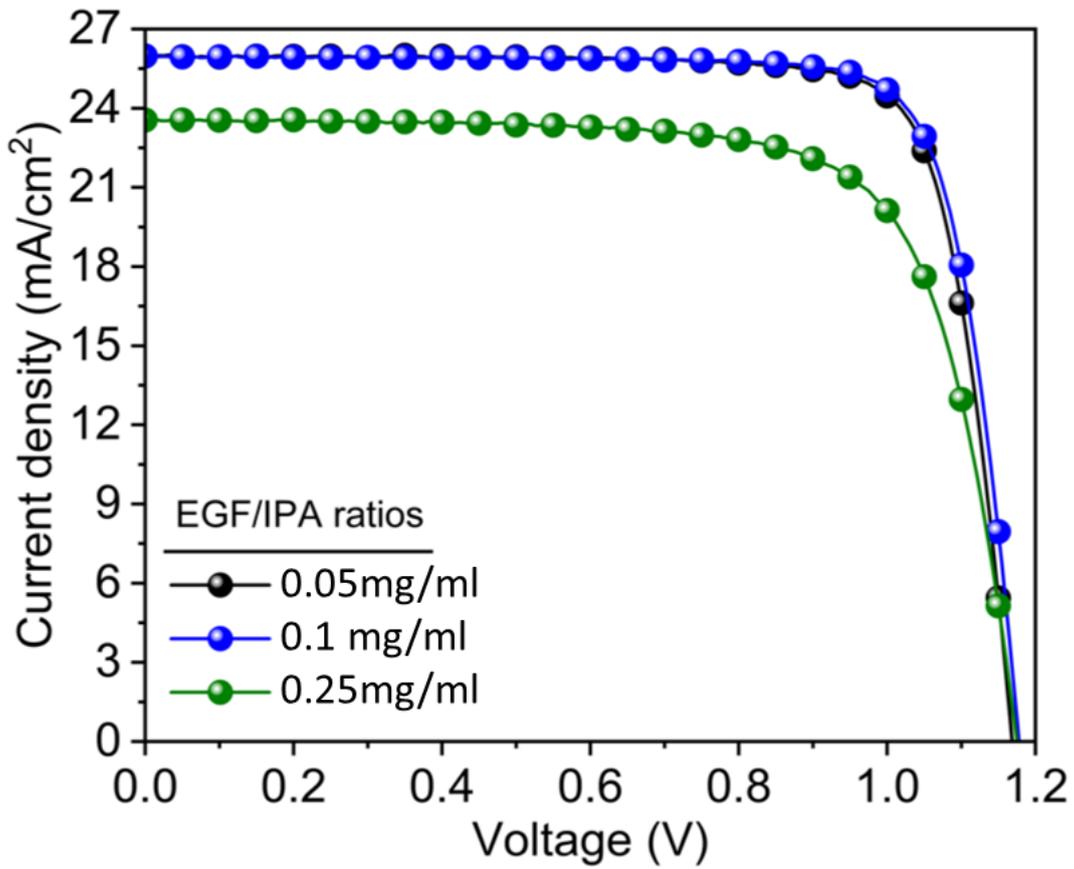

**Figure S3.** *J-V* characteristics curves using different EGF/ethanol ratio solutions.

**Table S1.** PV parameters extracted from Figure S6

| EGF to IPA ratio | $V_{oc}$(V) | $J_{sc}$ (mA/cm$^2$) | FF(%) | PCE(%) | |
|---|---|---|---|---|---|
| 0.05 mg/ml | 1.165 | 25.97 | 81.1 | 24.54 | |
| 0.1 mg/ml | 1.178 | 25.98 | 81.4 | 25.0 | Target |
| 0.25 mg/ml | 1.174 | 24.15 | 73.5 | 20.84 | |

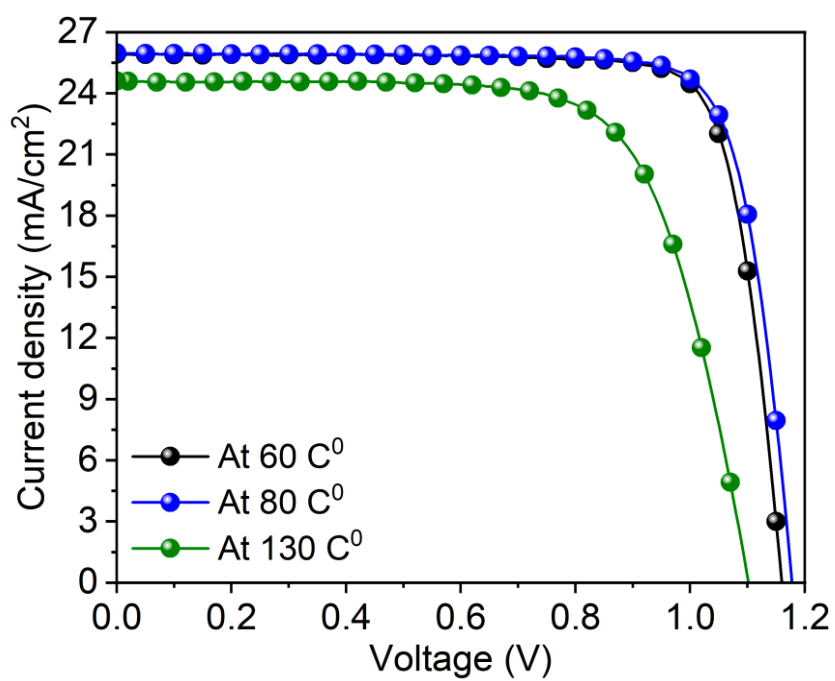

**Figure S4.** *J-V* characteristics curves at different preparation temperatures.

**Table S2.** PV parameters extracted from Figure S7

| Temperature | $V_{oc}$(V) | $J_{sc}$ (mA/cm$^2$) | FF(%) | PCE(%) | |
|---|---|---|---|---|---|
| 60 | 1.153 | 25.92 | 82.0 | 24.51 | |
| 80 | 1.178 | 25.98 | 81.4 | 25.0 | Target |
| 130 | 1.090 | 24.60 | 71.3 | 19.12 | |

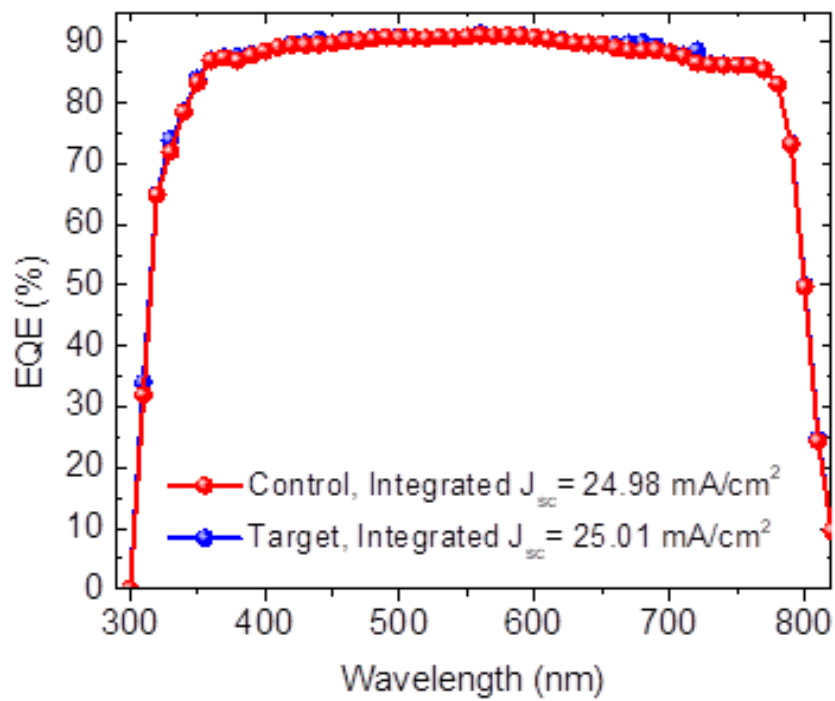

**Figure S5.** EQE spectra and photocurrent integrated over the standard AM 1.5G solar spectrum

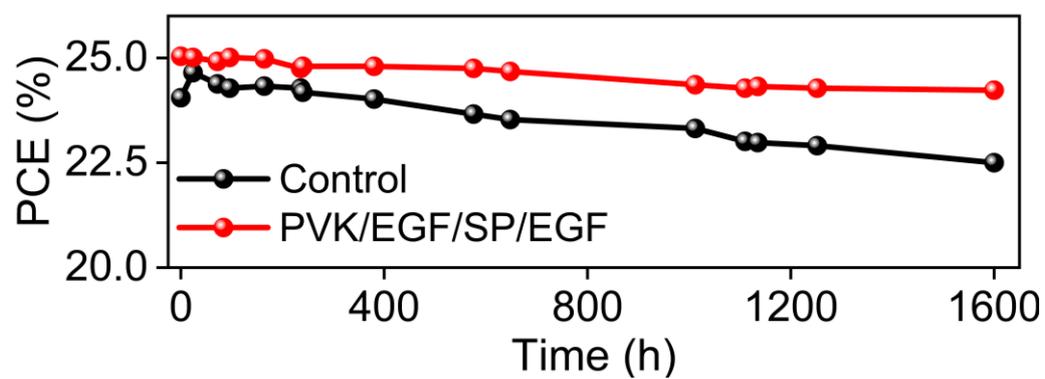

**Figure S6.** Shelf stability at RT and under ~8% RH of the control and target devices using Spiro-OMeTAD as HTL.

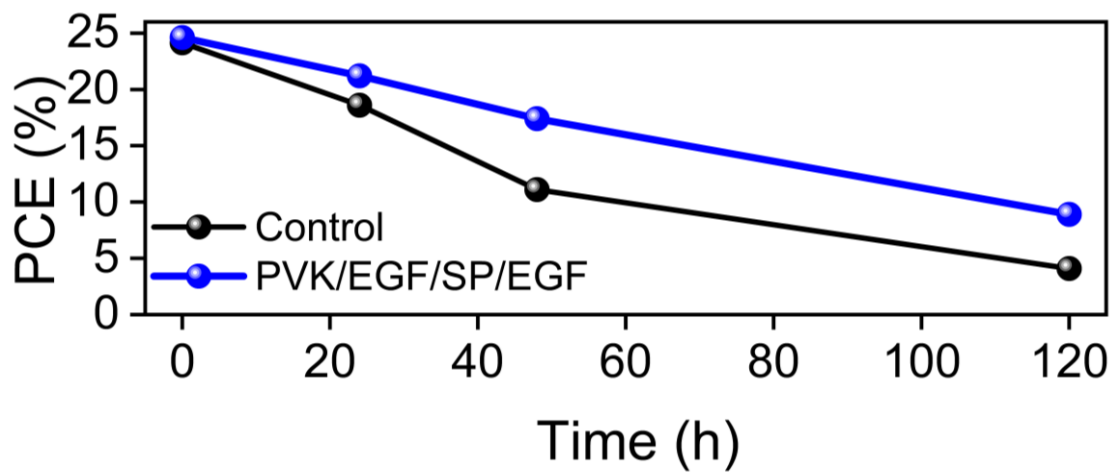

**Figure S7.** Shelf stability at 80 °C and ~75% RH of the control and target samples using Spiro-OMeTAD as the HTL.

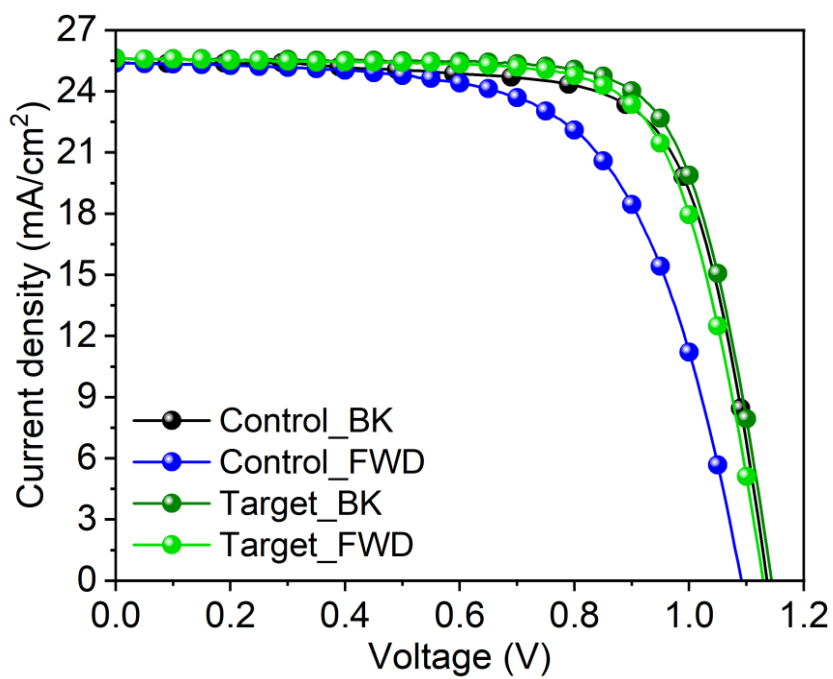

**Figure S8.** *J–V* results for devices using PTAA as HTL for control and target devices.

**Table S3.** PV parameters extracted from Figure S10

|  | $V_{oc}$(V) | $J_{sc}$ (mA/cm$^2$) | FF(%) | PCE(%) |  |
| --- | --- | --- | --- | --- | --- |
| Bk | 1.132 | 25.41 | 72.50 | 20.86 | control |
| FWD | 1.070 | 25.39 | 63.70 | 17.31 | control |
| Bk | 1.142 | 25.62 | 74.4 | 21.77 | Target |
| FWD | 1.121 | 25.61 | 73.3 | 21.04 | Target |

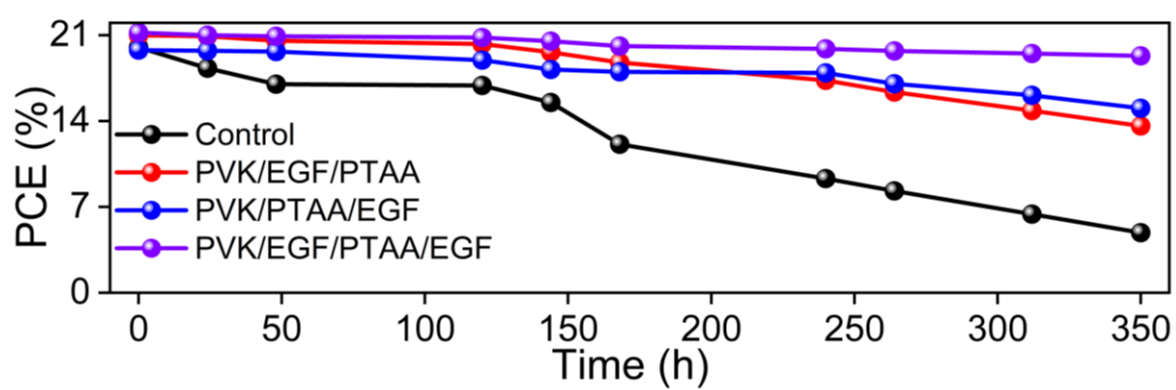

**Figure S9.** Shelf stability at RT and ~8% RH of control and target devices using PTAA as HTL.